\documentclass[10pt,twocolumn,letterpaper]{article}

\usepackage{iccv}
\usepackage{times}
\usepackage{epsfig}
\usepackage{graphicx}
\usepackage{amsmath}
\usepackage{amssymb}
\usepackage{multirow}

\usepackage{enumitem}

\usepackage[pagebackref=true,breaklinks=true,letterpaper=true,colorlinks,bookmarks=false]{hyperref}

\iccvfinalcopy 

\def\iccvPaperID{12706} 

\ificcvfinal\fi

\begin{document}

\title{RTNH+: Enhanced 4D Radar Object Detection Network using Combined CFAR-based Two-level Preprocessing and Vertical Encoding}

\author{Seung-Hyun Kong\thanks{co-first authors} \quad\quad Dong-Hee Paek\footnotemark[1] \quad\quad Sangjae Cho\\
CCS Graduate School of Mobility\\
KAIST, Daejeon, Korea\\ 
{\tt\small \{skong, donghee.paek, sanje\}@kaist.ac.kr}
}


\maketitle
\ificcvfinal\thispagestyle{empty}\fi

\begin{abstract}
Four-dimensional (4D) Radar is a useful sensor for 3D object detection and the relative radial speed estimation of surrounding objects under various weather conditions. However, since Radar measurements are corrupted with invalid components such as noise, interference, and clutter, it is necessary to employ a preprocessing algorithm before the 3D object detection with neural networks. In this paper, we propose RTNH+ that is an enhanced version of RTNH, a 4D Radar object detection network, by two novel algorithms. The first algorithm is the combined constant false alarm rate (CFAR)-based two-level preprocessing (CCTP) algorithm that generates two filtered measurements of different characteristics using the same 4D Radar measurements, which can enrich the information of the input to the 4D Radar object detection network. The second is the vertical encoding (VE) algorithm that effectively encodes vertical features of the road objects from the CCTP outputs. We provide details of the RTNH+, and demonstrate that RTNH+ achieves significant performance improvement of 10.14\% in ${{AP}_{3D}^{IoU=0.3}}$ and 16.12\% in ${{AP}_{3D}^{IoU=0.5}}$ over RTNH.
\end{abstract}

\section{Introduction}

The 4D Radar is a highly useful sensor for autonomous vehicles that need to detect the positions and movements of surrounding objects, as it can measure the range, azimuth, elevation, and Doppler frequency of the objects \cite{mimo}. A recent study \cite{paek1} indicates that 4D Radar shows much greater robustness and accuracy in rainy and snowy conditions than LiDAR. Furthermore, since the hardware performance of 4D Radar continues to improve, 3D object detection networks for 4D Radar are expected to be used widely. However, despite of the huge potential of 4D Radar in the autonomous driving, studies on the 4D Radar are still in the early stage and in this paper we identify two major directions to enhance 3D object detection with the 4D Radar.

RTNH \cite{paek1} is the first 3D object detection network for 4D Radar trained on a large dataset (a.k.a. K-Radar). The preprocessing algorithm of the RTNH selects only the measurements within top-10\% in power from the raw measurements of the 4D Radar. In a recent study \cite{paek2}, it is demonstrated that bird-eye-view (BEV) object detection and 3D object detection with RTNH using only the measurements within top-3\% and top-5\% in power results in better performance than RTNH \cite{paek1}. However, since the power of measurements for objects at a distance $r$ is proportional to ${r^{-4}}$, this preprocessing method can cause a huge loss of measurements for long-range objects, resulting in a poor detection performance.

In contrast, conventional and widely used preprocessing methods for Radar, such as Cell Averaging CFAR (CA-CFAR) and Ordered Statistic CFAR (OS-CFAR) determine their threshold adaptively for each local area instead of uniform threshold to the entire measurements of the Radar. However, since the vehicle environment is heterogeneous, where invalid measurements caused by noise, interference, and clutter can occur, the performance of CA-CFAR deteriorates significantly. Moreover, even CA-CFAR or OS-CFAR can cause a non-negligible loss of valid measurements for objects as a result of thresholding when there are many surrounding objects. Therefore, when using CA-CFAR or OS-CFAR with a low false alarm rate, we can remove a significant portion of invalid measurements, but this may increase miss-detection probability as a consequence. Furthermore, the sorting function in OS-CFAR requires a considerable amount of computational resources for a large data. This observation testifies the importance of efficient 4D Radar measurement preprocessing algorithm to improve the 3D object detection performance. In this paper, we propose a novel combined CFAR-based two-level preprocessing (CCTP) algorithm that generates measurements of different characteristics at two different filtering levels using the same measurements. As a result, the CCTP algorithm can enrich the information of the input to the 4D Radar object detection networks and significantly improves 3D object detection performance.

Another identified direction is the improvement of the object detection performance and real-time capabilities of deep neural networks with preprocessed 4D Radar measurements. RTNH \cite{paek1} and VoxelNet \cite{zhou} are neural networks for Radar and LiDAR, respectively, but both networks compress height information of 3D point clouds into a single value to project them onto a 2D BEV image in order to reduce the computational cost required for bounding box proposal. However, since height information is crucial for the recognition and localization of 3D objects, compressing height information into a single value can cause a degradation in 3D object detection performance. On the contrary, Yan et al. \cite{yan} propose a method to concatenate features along the height axis (i.e., vertical features) and features along the channel axis (i.e., channel features) to encode a greater amount of height information, resulting in a significant performance improvement (about 7.44\%) in 3D object detection. However, this approach requires a large amount of computational resources and memory. Therefore, in order to enhance the object detection performance of a 4D Radar neural network, it is necessary to encode vertical information effectively while reducing computational and memory requirements. In this paper, we propose a novel and efficient algorithm for encoding vertical features, called Vertical Encoding (VE). VE algorithm extracts important vertical features for 3D object detection without loss by learning vertical query that has high correlation with features along the vertical axis. Furthermore, unlike previous studies \cite{yan,shi}, VE algorithm requires only operations related to the height axis, resulting in a small computational and memory requirements with strong performance improvement.

In a summary, our contribution in this paper includes;

\begin{itemize}[noitemsep]
\item Combined CFAR-based two-level preprocessing (CCTP) algorithm to generate two filtered measurements of different characteristics using the same 4D Radar measurements, which significantly improves 3D object detection performance.
\item Vertical Encoding (VE) algorithm to appropriately encode vertical features of the objects with the CCTP algorithm output.
\item RTNH+ that is an enhanced 4D Radar object detection network based on RTNH integrated with novel CCTP and VE algorithms. We demonstrate that RTNH+ achieves 10.14\% ${AP}_{3D}^{IoU=0.3}$ and 16.12\% ${AP}_{3D}^{IoU=0.5}$ higher than RTNH on the K-Radar dataset.
\end{itemize}
In addition, we find that there are objects invisible to 4D Radar that is implemented on the front bumper of vehicles but visible to the LiDAR at the rooftop of the vehicles. This causes false labels in the 4D Radar data, when we use auto-labeling using LiDAR. In this paper, we provide correct revision to the 4D Radar labels in the K-Radar dataset.

This paper is organized as follows. In Section 2, we describe conventional Radar preprocessing algorithms related to our research and the current state-of-the-art 4D Radar object detection network, RTNH. In Section 3, we introduce the CCTP and VE algorithms proposed in this study, and RTNH+ implemented with these two proposed algorithms. In addition, we describe the label revision for objects invisible to Radar but visible to LiDAR applied to the K-Radar dataset. In Section 4, we present the performance evaluation results of the CCTP algorithm and RTNH+ with the CCTP and VE algorithms on the K-Radar dataset, and finally, we draw a conclusion of this paper in Section 5.

\section{Related Works}

In this section, we introduce existing studies related to the algorithms presented in this paper.

\subsection{Preprocessing algorithms for Radar measurements}

In general, Radar systems employ a preprocessing algorithm to remove invalid measurements from its raw measurements and extract valid measurements (i.e., signal measurements reflected from objects). This process involves calculating a threshold based on the statistical distribution of valid and invalid measurements and removing measurements with (signal) power below the threshold. The CFAR method is commonly used, with an assumption that the received noise signal follows a complex Gaussian distribution, and calculates a threshold for the measurement power corresponding to the maximum allowable false alarm rate (typically within 1\%) \cite{gandhi}.

However, the signal power distribution of components other than valid measurements can be heterogeneous, where measurements are corrupted by noise, interference signals from other surrounding objects, clutter around the target object. Therefore, the threshold for selecting valid measurements can vary in each local area within the whole detection area of the Radar. Despite this adaptive thresholding, CA-CFAR performs poorly in heterogeneous environments \cite{gandhi}, and its performance deteriorates significantly when multiple objects are present. OS-CFAR, which is often used when multiple objects are present in heterogeneous environments, requires a high computational cost for sorting and performs poorly in the presence of too many objects \cite{srinivasan}.

The conventional CFAR-based methods in Radar, as described above, has a single-step detection approach where the presence or absence of an object is determined from the thresholding results. Therefore, maintaining a low false alarm rate using CFAR in Radar is a critical factor for reliable detection and object tracking performance, but setting the threshold to a low false alarm rate can increase miss-detection probability to a large value.

\subsection{4D Radar Object Detection Network, RTNH, and Datasets}

Recently, large datasets collected by sensors such as RGB cameras, LiDAR, 3D, and 4D Radars for autonomous driving have become publicly available \cite{caesar,geiger,paek1,mao}, and deep neural network-based object detection \cite{lang,paek1} research utilizing these datasets is currently underway. Among these sensors, 4D Radar has recently become commercially available, and there are only a few released datasets \cite{paek1,palffy,zheng,meyer}.

Astyx \cite{meyer} is the first publicly available 4D Radar point cloud (4DRPC) dataset, and Xu et. al \cite{xu} demonstrate the capability of 4D Radar for object detection using a PointPillars \cite{lang}-based 3D object detection network trained on 4DRPC. However, Astyx is a small dataset (0.5K), making it difficult to develop a high-performing 4D Radar object detection network. VoD \cite{palffy} and TJ4DRadSet \cite{zheng} have provided more frames (8.7K and 7.8K, respectively) of 4DRPC data and attempted 3D object detection using PointPillars \cite{lang}. However, as these datasets are acquired in clear weather conditions in urban environments, these datasets are still not enough to train a neural network for 3D object detection in various weather conditions. On the other hand, K-Radar \cite{paek1}, acquired in severe weather conditions (e.g., fog, rain, and snow) and various road environments (e.g., urban, suburban, alleys, highways), provides a large (35K) 4D Radar raw measurement (i.e., 4D Radar tensor a.k.a. 4DRT) dataset, and Paek et. al \cite{paek1} demonstrate improved 3D object detection performance using RTNH (Radar Tensor Network with Height) trained on K-Radar. However, RTNH uses a simple preprocessing algorithm that extracts only measurements within top-10\% in power and compresses various height-related information to a single value for object detection on BEV feature maps, which can not fully utilize vertical information of objects.

\section{Proposed Algorithm and Contribution}

RTNH has a 3-stage architecture composed of measurement preprocessing, 3D sparse backbone (3DSB), and neck \& head (N\&H). In this section, we propose two novel algorithms to enhance RTNH, the combined CFAR-based two-level preprocessing (CCTP) algorithm and the vertical encoding (VE) algorithm, which are related to the 1st and 2nd stages of RTNH, respectively.

\subsection{Combined CFAR-based Two-Level Preprocessing (CCTP) Algorithm}

A typical Radar detection algorithm based on CFAR guarantees a very low (less than 1\%) false alarm rate by considering measurements above a strict threshold as valid detections. Therefore, in CFAR, when the signal-to-noise ratio (SNR) of valid measurements is very low, most of the valid measurements can be removed, whereas when the SNR of valid measurements is high, most of the noisy measurements can be removed, resulting in a desirable outcome. In contrast, the preprocessing method of RTNH extracts and uses the top-5\% or top-3\% of measurements in power regardless of SNR \cite{paek2}. Thus, this method requires a high amount of computational cost for sorting, and can lead to a high miss-detection rate due to the loss of measurements for distant objects.

\begin{figure}[b!]
\centering
\includegraphics[width=1.0\columnwidth]{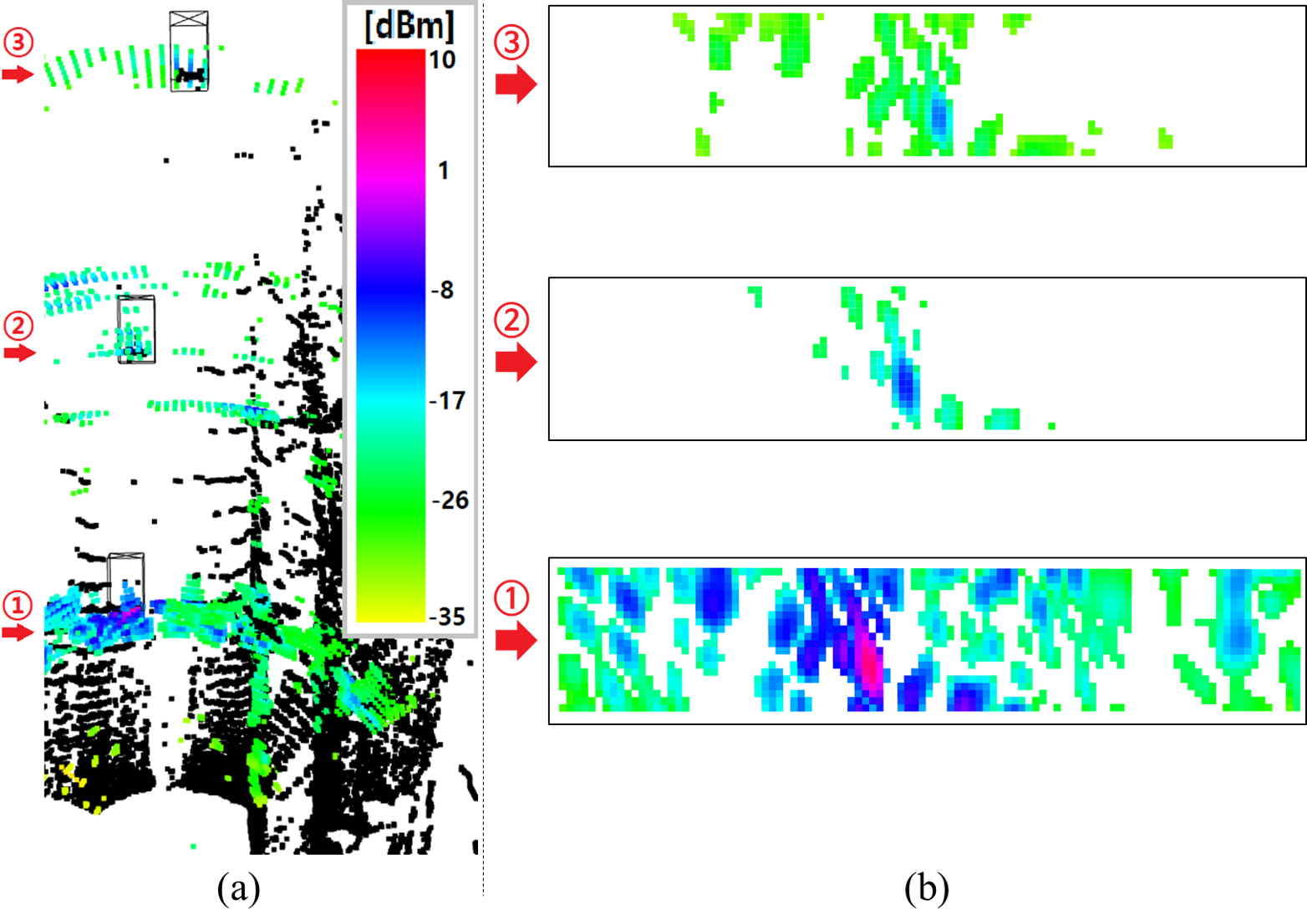}
\caption{Example of 4D Radar measurements with CA-CFAR 0.05: (a) 4D Radar BEV plot with colored signal power and 3 bounding boxes over LiDAR point cloud in black, (b) Azimuth-Height plots (front view) of the measurements at particular ranges indicated by red arrows on the left side of (a).}
\label{fig:pw}
\end{figure}

As illustrated in Fig.\ref{fig:pw}, it can be observed that both the number and power of measurements decrease as the distance to the objects increases. Fig.\ref{fig:pw}-(b) displays the measurements and their (signal) power within the azimuth range of [-45$^{\circ}$, 45$^{\circ}$] and the elevation range of [-0.5m, 4m] for three ranges where objects are present as shown in Fig.\ref{fig:pw}-(a). As shown in Fig.\ref{fig:pw}-(b), the measurement with the highest power appears at the location of the object, and as the azimuth increases or decreases from the object location, the intensity of the measurements gradually decreases. This is because when FFT results contain invalid components such as sidelobe, noise, interference, and clutter, those invalid components are not easily removed with a CFAR not low enough. Moreover, in adverse weather conditions such as heavy rain, SNR of Radar measurements becomes low, and there is no significant difference between the power distributions of valid and invalid measurements. Therefore, the existing preprocessing methods with a low CFAR (e.g., 1\% or less) may limit false alarms to a low level but can increase miss-detection probability to a high level. On the other hand, using a much higher CFAR than 1\% can extract reliable and valid measurements, but can preserve a significant number of invalid measurements that can cause false alarms.

\begin{figure*}[ht!]
\centering
\includegraphics[width=2.0\columnwidth]{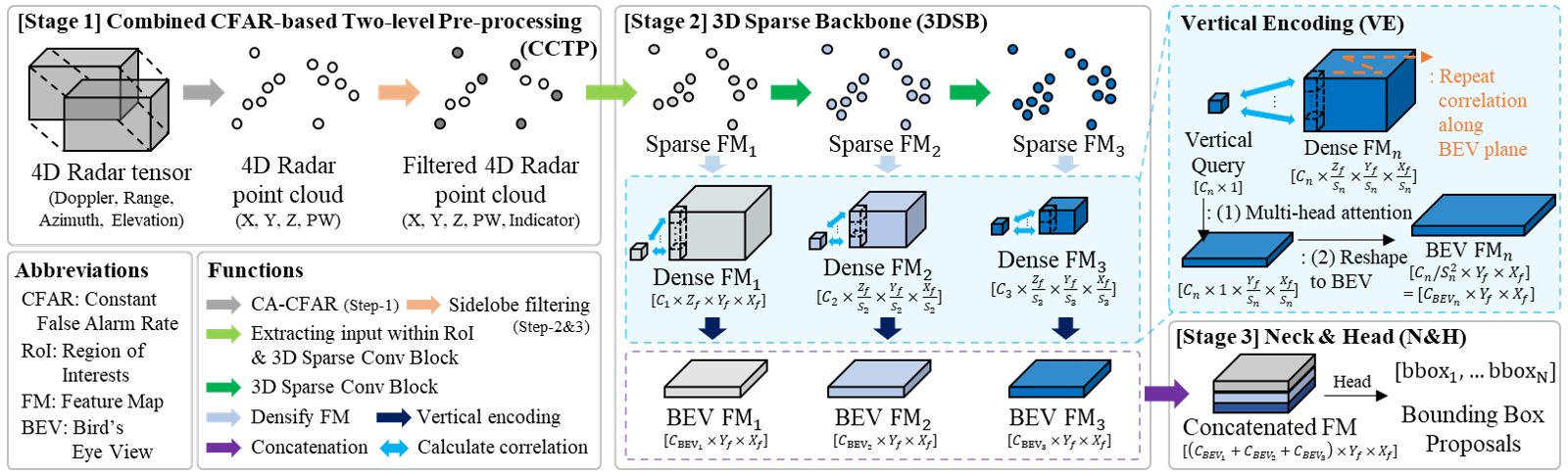}
\caption{The overall structure of RTNH+, where $C_n$, $C_{BEV_{n}}$, and $S_n$ represent the channel size of the $n$-th stage FM and BEV FM, and stride size, respectively, and $X_f$, $Y_f$, and $Z_f$ represent the size of the X, Y, and Z axes of the FM at the stage-1.}
\label{fig:net}
\end{figure*}

The proposed CCTP algorithm applies coarse CA-CFAR with a relatively high threshold to the incoming 4D Radar measurements in order to output as many valid measurements (related to objects) as possible without allowing too many invalid measurements. In the second step, it applies OS-CFAR with a lower threshold than the previous step to the preserved measurements (along azimuth axis) in order to minimize invalid measurements and output reliable and valid measurements. Notice that the environment is heterogeneous. However, in the first step, coarse CA-CFAR does not loose valid measurements due to the high enough threshold, and in the second step, OS-CFAR with low threshold is robust to heterogeneous environments and is applied to measurements along azimuth axis that has the same attenuation. This approach can extract two types of measurements with different characteristics using the same 4D Radar measurements, providing more information useful for object detection network.

In detail, in the first step (CCTP step-1) of the CCTP algorithm, a coarse CA-CFAR with a higher threshold ${K_1}$\% ${(0\leq{K_1}\leq100)}$ than the conventional CFAR (around 1\%) is applied to all measurements to remove noise measurements with very low power. Since the Radar measurements have a large power attenuation with respect to the distance, in CCTP step-2, OS-CFAR with a low threshold of ${K_2}$\% ${(0\leq{K_2}\leq100)}$ is applied to measurements along azimuth axis of the same range (i.e., same attenuation). OS-CFAR is preferred over CA-CFAR because multiple objects can be present along the azimuth axis for a given range, and it is suitable for heterogeneous environments. To increase the sorting operation speed required for order statistics, we compress the vertical (elevation) axis to a single value to produce an azimuth-wise 1D vector, to which OS-CFAR is applied. In other words, for each range index ${i_r (0\leq{i_r}<N_r)}$, the first processing of CCTP step-2 compresses the measurements on a 2D space (of azimuth and elevation) using a vertically-weighted sum to produce an azimuth-wise 1D vector. Let $M_1^{2D}[i_r, :, :]$ be measurements expressed on a 2D space (by azimuth and elevation axes) for a range index $i_r$. Since the objects on the road are located relatively at low elevation above the ground as shown in Fig.\ref{fig:pw}-(b), we project $M_1^{2D}[i_r,:, :]$ onto the range-azimuth plane after a vertically-weighted sum. When we denote  $W_V$ as the vertical weight vector, the resulting weighted projection (1D vector) becomes
\begin{align}
    \nonumber
    {{M}^{1D}_{1}}[{{i}_{r}},:]&=\sum_{{{i}_{e}}=0}^{{N}_e-1}{{W}_{V}}[{{i}_{e}}]{{M}_{1}}[{i_r},:,{i_e}]\\ 
    &=\sum_{i_e=0}^{N_e-1}(E_T-i_e)M_1[i_r,:,i_e],
\end{align}
where $i_a$ ${(0\leq{i_a}<{N_a})}$ and $i_e$ ${(0\leq{i_e}<{N_e})}$ are the azimuth and elevation indices, respectively, $E_T$ is the size of the $M_1$ in elevation axis,
\begin{align}
    {{W}_V}[i_e]=E_T-i_e,
\end{align}
and $i_e$=0 indicates the lowest index of the elevation axis. We apply OS-CFAR to $M_1^{1D}[i_r,:]$ (1) to find azimuth indices where the elements of $M_1^{1D}[i_r,:]$ is within top-$K_2$\% in power among the elements of $M_1^{1D}[i_r,:]$. Note that OS-CFAR applied to the $M_1^{1D}[i_r,:]$ of size [1$\times$${N_a}$] requires $O({N_a}\log({N_a}))$ for sorting process, which is $N_e$ times lower computation than RTNH, where sorting is applied to the entire measurements $M_1^{3D}[:,:,:]$ to require $O({N_r}{N_a}{N_e}\log({N_r}{N_a}{N_e}))$ computations. When CCTP step-2 is applied to all of the range indices (${i_r}$=0, 1, …,$N_r$-1), we can find preserved measurement locations [$j_r$,$j_a$] on the range-azimuth plane, where ${j_r}\in{J_r}$, ${j_a}\in{J_a}$, ${J_r}\subset\{0,1,…,{N_r}-1\}$, and $J_a\subset\{0,1,…,{N_a}-1\}$. Finally, we find the 3D output $M^{3D}_2$ of the CCTP step-2 as 
\begin{align}
    M^{3D}_2[j_r,j_a,:]=
    \begin{cases}
    M_1^{3D}[j_r,j_a,:],&\text{for condition A} \\
    0, &\text{otherwise},
    \end{cases}
\end{align}
where condition A is `${j_r}\in{J_r}$ and ${j_a}\in{J_a}$'. Since the 3D output $M_2^{3D}$ of the CCTP step-2 is a result of OS-CFAR applied to azimuth-wise 1D vector for each range independently, there could be a valid but removed measurements next to the preserved (i.e., survived) measurements along the range axis. In fact, this may be possible along the azimuth axis, too, for a small $K_2$. Therefore, in CCTP step-3, we recover removed but possibly valid measurements (in the CCTP step-2) next to the preserved measurements $M_2^{3D}$. To this purpose, we set allowed distance $d_r$ and $d_a$ in the range and azimuth axes, respectively, and find the final output $M_3^{3D}$ of the CCTP algorithm (step-3) as
\begin{align}
    {M_3^{3D}[i_r,i_a,:]}=
    \begin{cases}
    {M_1^{3D}[i_r,i_a,:]}, &\text{for condition B} \\
    0, &\text{otherwise},
    \end{cases}
\end{align}
where $M_1^{3D}[:,:,:]$ is the raw 4D Radar measurements and condition B is `$[i_r-d_r,i_r+d_r]$ contains any ${j_r}\in{J_r}$ and $[i_a-d_a,i_a+d_a]$ contains any ${j_a}\in{J_a}$'. Note that since CCTP step-2 applies OS-CFAR for each range independently, there could be more valid-but-removed measurements along the range axis than azimuth axis so that we use ${d_r}>{d_a}\geq{1}$. However, since large $d_r$ and $d_a$ could result in an inclusion of many invalid measurements, we use $d_r$=2 and $d_a$=1 for a minimum recovery. 

In a summary, the proposed CCTP algorithm produces two types of measurements with different statistical characteristics in the two sequential filtering processes. These two types of measurements can enrich the informatiojn of the input to the neural network for the 3D object detection. In stage-2 of RTNH+, these two outputs are concatenated and appropriately utilized for 3D object detection. Specifically, the output of CCTP step-1 becomes the primary input for object detection as it contains a relatively large number of valid measurements, and the output of CCTP step-3 serves as a reliability indicator additional to the primary input to the RTNH+ stage-2.


\begin{figure*}[ht!]
\centering
\includegraphics[width=2.0\columnwidth]{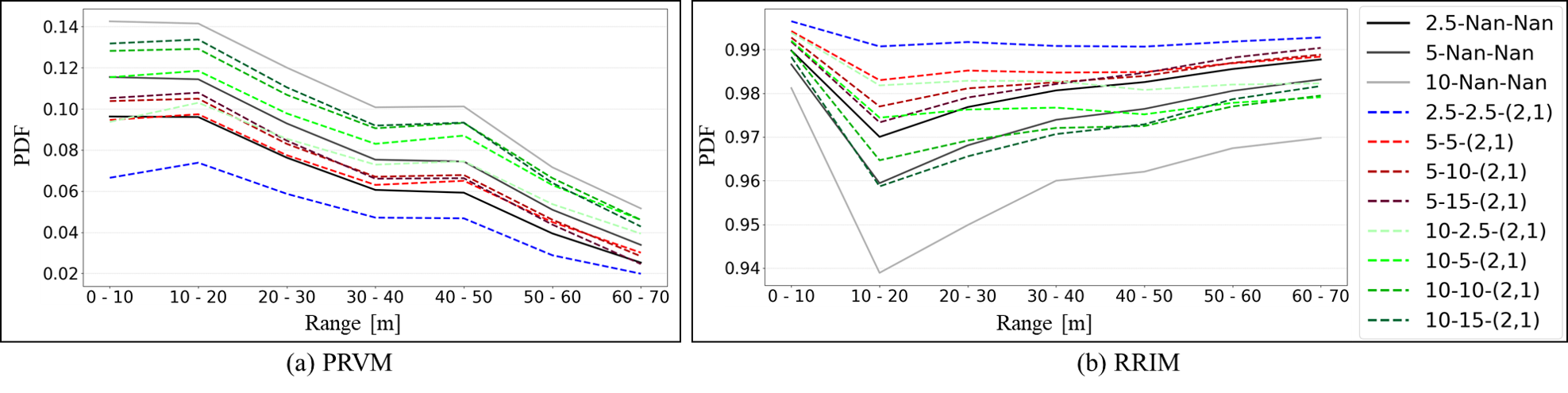}
\caption{Results of CCTP algorithm with various hyperparameters, where the legend shows $K_1$-$K_2$-($d_r$, $d_a$). Therefore, ``$K_1$-Nan-Nan'' represents the performance of the CCTP step-1 output, and ``$K_1$-$K_2$-Nan'' shows the performance of the CCTP step-2 output. (a) PRVM, (b) RRIM.}
\label{fig:plot}
\end{figure*}

\subsection{Enhanced 4D Radar Object Detection Network using Vertical Encoding}

The proposed RTNH+ consists of three stages (i.e., CCTP, 3DSB, and N\&H) similar to RTNH, as shown in Fig.\ref{fig:net}. However, unlike RTNH that (1) uses top-10\% strong measurements and (2) encodes height information to a single value, RTNH+ (1) uses the CCTP algorithm to extract two filtered measurements with low computational cost and (2) encodes useful vertical features for 3D object detection through correlation with vertical query.

The first stage of RTNH+ (i.e., stage-1)  is the 3-step CCTP algorithm that produces input to the 3DSB from the 4D Radar tensor. The output of CCTP step-1 (i.e., CA-CFAR output) is used as the primary input for 3DSB, while the final output of CCTP step-3 is used as a complimentary indicator, that has a value of 1 for valid measurements and 0 for removed measurements. These indicators can provide very effective and useful information for precise object localization. Subsection 4.3 demonstrates that using the output of CCTP step-3 as indicator can significantly improve the strict ${AP}_{BEV}^{IoU=0.5}$ over the previous RTNH.

In RTNH+ stage-2, the output of CCTP is encoded into a sparse feature map (Sparse FM) with multi-resolution using 3D sparse convolution \cite{liu}. The Sparse FM is densified to produce Dense FM and then further encoded into a BEV FM through vertical encoding (VE) algorithm that consists of two steps as shown in Fig.\ref{fig:net} (upper right). In VE step-1, vertical features (of size [${C_n}\times{Z_f/S_n}$]) along the height axis of the Dense FM are correlated with vertical query (of size [${C_n}\times{1}$]) for $Z_f/S_n$ times to produce a compressed vertical feature that is highly correlated with vertical query. To reduce computational load in the VE step-1, multi-head attention (MHA) \cite{vaswani} is employed for the correlation, and the key hyper-parameters of MHA, such as the number of heads, are determined through ablation studies (refer to the Appendix). Therefore, the proposed VE in RTNH+ is different from the height encoding in RTNH \cite{paek1}, where vertical features are compressed into a single value and thus significant information about 3D objects can be lost. The proposed VE is also different from those in \cite{yan,shi}, where high computational load and memory are required to encode much more vertical features from channel features concatenated with vertical features. The proposed VE significantly improves 3D object detection performance by encoding important vertical features (e.g., the top and bottom surfaces and boundaries of 3D bounding boxes) with much lower computation and memory than \cite{yan,shi}. In VE step-2, the output of MHA is reshaped to fit to the final size of BEV FM (i.e., [${Y_f}\times{X_f}$]), which is to encode high-resolution features into deep-level features as demonstrated in previous research \cite{paek3}.

The stage-3 Neck \& Head of RTNH+ is identical to that of RTNH (refer to Appendix), where we extract 3D bounding boxes from the fused multi-resolution BEV FM.

\subsection{Additional Contribution}

As mentioned in subsection 2.2, most available 4D Radar object detection datasets use LiDAR-detected 3D bounding boxes as reference labels. However, while LiDAR is installed on the rooftop of a vehicle, 4D Radar is installed in front of or directly above the vehicle bumper. Due to this difference in the sensor placement, objects visible to LiDAR may not be visible to 4D Radar. This problem results in the existence of object labels that are actually invisible to Radar in 4D Radar datasets. For example, in K-Radar, labels of objects invisible to 4D Radar but visible to LiDAR account for 25.58\% of all labels, which can negatively affect the performance of the 4D Radar object detection neural network and cause a high level of false alarms in actual inference. Therefore, we demonstrate that distinguishing labels invisible to Radar can improve the performance of 4D Radar object detection neural networks, and we provide revised labels for K-Radar. 

\begin{figure*}[t!]
\centering
\includegraphics[width=2.0\columnwidth]{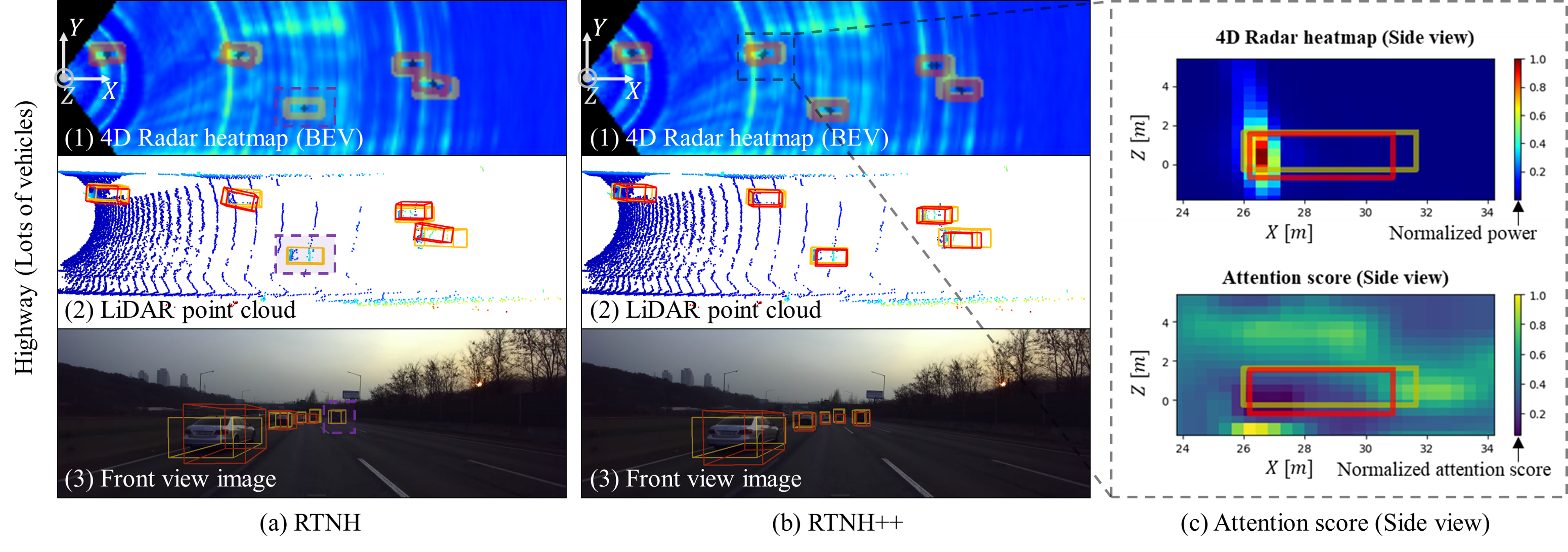}
\caption{An example of 3D object detection using (a) RTNH and (b) RTNH+, in a highway environment. (a) and (b) show that (1) 4D Radar heatmap in BEV, (2) Lidar point cloud, and (3) front view image with bounding boxes, at the same site. Bounding boxes with yellow solid line, red solid line, and purple dashed line represent the ground truths, predictions, and miss detections, respectively. (c) shows the 4D Radar heatmap of the object and the attention score of vertical queries on a side-view (on the XZ-plane). Additional examples are available in Appendix.}
\label{fig:result}
\end{figure*}

\section{Experiments}

In this section, we use K-Radar dataset to evaluate the performance of the proposed CCTP and VE algorithms and the performance of the RTNH+ with the two proposed algorithms. For a comparison, we evaluate the performance of RTNH and various different implementations of CCTP and VE to the RTNH+. 

\subsection{Experimental Setup and Metric}

\noindent\textbf{Implementation Detail} We implement RTNH+ using PyTorch 1.11.0 on a Ubuntu machine equipped with an RTX3090 GPU. We set the batch size and learning rate of RTNH+ to 4 and 0.0005, respectively, and trained RTNH+ using the AdamW optimizer \cite{loshchilov} for 11 epochs. For training and performance evaluation, we used K-Radar \cite{paek1}. To compare RTNH+ with its baseline network, RTNH, we set the same region of interest (RoI) size and sedan class as the detection target, where RoI is within 0$\sim$72[m] in range, -6.4$\sim$6.4[m] in width, and -2$\sim$6[m] in height.

\noindent\textbf{Metric} We analyze the preserved rate of valid measurements (PRVM) and removed rate of invalid measurements (RRIM) of the CCTP algorithm and propose hyperparameters of the CCTP appropriate for RTNH+. We evaluate the object detection performance of RTNH+ using the Intersection over Union (IoU)-based Average Precision (AP) metric. We show AP performance in both BEV (${AP}_{BEV}$) and 3D (${AP}_{3D}$), as in \cite{geiger}, and demonstrate the accurate localization performance of RTNH+ using the AP performance for IoU=0.3 and 0.5.

\subsection{CCTP Algorithm}

To evaluate the performance of the proposed CCTP algorithm, we investigate the PRVM within objects’ bounding boxes and the RRIM outside the bounding boxes. In the following subsection, we compare the improvement of the final object detection performance of RTNH+ with the CCTP algorithm.

Fig.\ref{fig:plot} presents the PRVM and RRIM across multiple range intervals for various CCTP algorithms evaluated on the K-Radar dataset. Recent study \cite{paek2} on hyperparameters in RTNH shows that using measurements of top-3\% and top-5\% in power results in higher performance in both 3D and BEV object detection, respectively, compared to the cases of using measurements of top-10\% in power. In addition, we exclude cases for $K_1>10$ from the following experiments, because the input size for stage-2 of RTNH+ becomes large and the computation cost significantly increases. 

As shown in Fig.\ref{fig:plot}, PRVM and RRIM of various preprocessing algorithms exhibit opposite trends, in other words, when PRVM is higher, RRIM becomes lower. However, it should be noticed that small drop in PRVM may cause some (not large) drop in the detection probability of the RTNH+, but small drop in PRIM can cause a huge increase in the false alarm probability of the RTNH+. This is because the number of invalid measurements is far much larger than the number of valid measurements in usual 4D Radar measurements, which is also observed in K-Radar. Therefore, while good performance in PRVM is desired, performance in RRIM should be kept to a high level to improve the overall performance of RTNH+. To accurately evaluate the performance of the various preprocessing algorithms in Fig.\ref{fig:plot}, we need to evaluate 3D object detection performance using the resulting measurements of each preprocessing algorithm. However, we can find some observation on the results in Fig.\ref{fig:plot}, where the case “2.5-2.5-(2,1)” shows the best (among the plotted graphs) performance in RRIM but worst (too low) in PRVM, and the case “10-$K_2$-(2,1) with $K_2>0$” shows good performance in PRVM but poor in RRIM. On the other hand, the case “5-$K_2$-(2,1) with $K_2>0$” shows good performance in RRIM and moderate performance in PRVM. Since $K_2=5$ has best performance in PRIM among various $K_2$ values, we may expect that, among the variations in Fig.\ref{fig:plot}, the case “5-5-(2,1)” can lead to the best performance of RTNH+, which is testified in the next subsection and agrees with the results introduced in \cite{paek2}. Note that $K_1$ and $K_2$ slightly different from 5 maybe useful, however, in the experiments in the next subsection, the cases evaluated in Fig.\ref{fig:plot} are fine enough to approximate the best values of $K_1$ and $K_2$.

\subsection{Comparison of RNTH++ to RTNH}

The following Table \ref{tab:rtnh} and Fig.\ref{fig:result} demonstrate the quantitative and qualitative 3D object detection performance of RTNH and RTNH+ on K-Radar dataset. As shown in Table \ref{tab:rtnh}, RTNH+ shows (1) higher 3D object detection performance, since it has both lower miss-detection and lower false alarm by appropriately utilizing the CCTP step-1 output and the CCTP step-3 output, respectively, and (2) more precise localization performance by utilizing the indicator information from the CCTP step-3 output. The result in Table \ref{tab:rtnh} indicates that RTNH+ achieves 10.14\% and 7.3\% increase in ${AP}_{3D}^{IoU=0.3}$ and ${AP}_{BEV}^{IoU=0.3}$, respectively, over the RTNH \cite{paek1}, and the following explains this performance improvement of 3D object detection in detail.

\begin{table}[ht!]
\begin{center}
\begin{tabular}{c|cc|cc}
\hline\hline
\multirow{2}{*}{Network} & \multicolumn{2}{c|}{IoU=0.3} & \multicolumn{2}{c}{IoU=0.5} \\
\cline{2-5}
      & ${AP}_{3D}$  & ${AP}_{BEV}$ & ${AP}_{3D}$ & ${AP}_{BEV}$ \\
\hline
 RTNH \cite{paek1} & 47.44 & 58.39 & 15.60 & 43.18 \\
 RTNH \cite{paek2} & 47.93 & 59.37 & 16.51 & 45.44 \\
 RTNH+ & \textbf{57.58} & \textbf{65.69} & \textbf{31.72} & \textbf{54.66} \\
\hline
\hline
\end{tabular}
\end{center}
\caption{Performance comparison of RTNH and RTNH+ on K-Radar dataset.}
\label{tab:rtnh}
\end{table}

\begin{table*}[ht!]
\begin{center}
\begin{tabular}{c|c|ccc|ccc|cc|cc|c|cc}
\hline\hline
 \multirow{2}{*}{Net.} & \multirow{2}{*}{RL} & \multicolumn{3}{c|}{CCTP} & \multicolumn{3}{c|}{Vertical Features} & \multicolumn{2}{c|}{IoU=0.3} & \multicolumn{2}{c|}{IoU=0.3} & Mem & \multicolumn{2}{c}{FPS [Hz]}\\
\cline{3-12}\cline{14-15}
 & & s1 & s2 & s3 & SVE & VCCE & VE & ${AP}_{3D}$ & ${AP}_{BEV}$ & ${AP}_{3D}$ & ${AP}_{BEV}$ & [MB] & S1-3 & S2 \\
\hline
 \cite{paek1}   &  & 10\% &  &  & \checkmark &  &  & 47.4 & 58.4 & 15.6 & 43.2 & 421 & 15.2 & 51.5 \\
 $N_a$   & \checkmark & 10\% &  &  & \checkmark &  &  & 49.3 & 60.5 & 18.4 & 45.3 & 421 & 15.2 & 51.5 \\
 $N_b$   & \checkmark & \checkmark &  &  & \checkmark &  &  & 52.9 & 60.5 & 26.0 & 45.1 & \textbf{359} & \textbf{15.8} & 97.9 \\
 $N_c$   & \checkmark & \checkmark & \checkmark &  & \checkmark &  &  & 53.5 & 60.9 & 25.1 & 46.2 & 359 & 12.4 & 96.5 \\
 $N_d$   & \checkmark & \checkmark & \checkmark & \checkmark & \checkmark &  &  & 53.8 & 63.0 & 26.4 & 46.3 & 359 & 8.91 & 97.3 \\
 $N_e$   & \checkmark & \checkmark & \checkmark & \checkmark & & \checkmark &  & 55.7 & 63.8 & 28.9 & 48.6 & 769 & 8.68 & 91.3 \\
 ours   & \checkmark & \checkmark & \checkmark & \checkmark &  &  & \checkmark & \textbf{57.6} & \textbf{65.7} & \textbf{31.7} & \textbf{54.7} & 502 & 9.46 & \textbf{107.5} \\
\hline
\hline
\end{tabular}
\end{center}
\caption{Ablation study to assess effects of revised label (RL), CCTP and VE algorithms for RTNH+ and other variants of RTNH+. `Net.', `s1`$\sim$`s3', and `10\%' represent networks, CCTP step-1$\sim$3, and selecting top-10\% strong measurements, respectively. `SVE' and `VCCE' represent encoders to encode vertical features into a single value \cite{paek1,zhou} and to encode vertical features via concatenating channel and vertical features \cite{yan,shi}, respectively. `Mem' represent the maximum GPU memory. 'S1-3' and 'S2' represent the computation times for the whole stages and for the stage-2 only, respectively.}
\label{tab:ablation}
\end{table*}

First, comparing to RTNH \cite{paek1} and \cite{paek2}, where measurements with top-10\% and top-5\% in power, respectively, are used, RTNH+ employs CCTP algorithm to produce filtered measurements of different characteristics in two different filtering levels (i.e., CCTP step-1 and step-3 outputs). And one of the two filtered measurements mostly contains reliable and valid measurements that are used for an indicator effective to precise object localization. In addition, as shown in Table \ref{tab:rtnh}, RTNH+ has a highly improved performance in the strict BEV objection detection metric ${AP}_{BEV}^{IoU=0.5}$ by 11.48\% over RTNH.

Second, RTNH+ effectively encodes vertical features that are highly correlated with vertical query using the proposed VE, whereas RTNH compresses vertical features into a single value and loses useful vertical information for 3D object detection. As demonstrated in Fig.\ref{fig:result}-(c), vertical query intensively encodes the vertical boundaries of 3D bounding boxes more than the whole measured vertical values. This leads to a strong 16.12\% improvement of RTNH+ over the RTNH in the strict 3D object detection performance metric ${AP}_{3D}^{IoU=0.5}$.

\subsection{Ablation study}

In Table \ref{tab:ablation}, we present the performance evaluation and the maximum GPU memory and computational cost required for inference in units of MB and frames per second (FPS), respectively. In the comparison of computation time for VE with SVE and VCCE, we measure the computation time for the whole network (of 3 stages) and for the stage-2 only. Note that the various networks we compare in Table \ref{tab:ablation} is denoted by ${N_a}\sim{N_e}$.

\noindent\textbf{Effects of revised label} In Table \ref{tab:ablation}, we compare RTNH trained using the existing labels to RTNH ($N_a$) trained using the revised labels. As expected, $N_a$ shows noticeable improvements in ${AP}_{3D}^{IoU=0.3}$, ${AP}_{BEV}^{IoU=0.3}$, ${AP}_{3D}^{IoU=0.5}$, and ${AP}_{BEV}^{IoU=0.5}$ by 1.88\%, 2.1\%, 2.8\%, and 2.12\%, respectively. Therefore, we use revised labels to train RTNH+ and show its performance in Table \ref{tab:ablation}.

\noindent\textbf{Effects of CCTP} We compare the performance improvement using each step of CCTP while the other conditions are fixed. As shown with GPU memory and FPS Total in Table \ref{tab:ablation}, all of the neural networks, ${N_b}\sim{N_d}$, that employ different levels of CCTP use the same CCTP step-1 output, which results in almost the same GPU memory but different computational costs for using different levels of CCTP. However, as mentioned in subsection 4.3, CCTP step-2 and step-3 produce indicators for reliable measurements and, thus, improve object localization accuracy. For example, the neural network $N_d$ that employs the full CCTP three steps shows noticeable improvement in ${AP}_{3D}^{IoU=0.3}$, ${AP}_{BEV}^{IoU=0.3}$, ${AP}_{3D}^{IoU=0.5}$, and ${AP}_{BEV}^{IoU=0.5}$ by 4.51\%, 2.53\%, 7.96\%, and 1.01\%, respectively, over the neural network $N_a$ that just uses top 10\% strong measurements.

\noindent\textbf{Effects of VE} We compare the performance of the VE (algorithm) to those of SVE and VCCE, while other conditions (e.g., CCTP) are fixed. From the experiments, we observe that there are three advantages in RTNH+ over other networks $N_d$ and $N_e$ using SVE and VCCE, respectively.
First, RTNH+ achieves strong improvement in ${AP}_{3D}^{IoU=0.3}$, ${AP}_{BEV}^{IoU=0.3}$, ${AP}_{3D}^{IoU=0.5}$, and ${AP}_{BEV}^{IoU=0.5}$ over (1) $N_d$ and (2) $N_e$ by (1) 3.75\%, 2.67\%, 5.3\%, and 8.35\% and (2) 1.89\%, 1.9\%, 2.86\%, and 6.05\%, respectively. This is because RTNH+ effectively encodes vertical features that is highly correlated with vertical query using the proposed VE, whereas $N_d$ and $N_e$ are not effectively encoding vertical features. 
Second, RTNH+ uses 34.7\% less GPU memory than $N_e$. This shows that VE effectively encodes vertical features with much lower memory cost than VCCE.
Third, RTNH+ achieves fast inference (i.e., high FPS) when compared to $N_d$ and $N_e$. As discussed in subsection 3.2, this is because (1) VE employs computationally efficient multi-head attention (MHA), and (2) when resizing BEV FM for concatenation, VE uses reshaping that does not require additional computation, whereas SVE and VCCE use convolution functions.

\section{Conclusion}

In this paper, we have proposed RTNH+, a significant enhancement of the baseline 4D Radar object detection network RTNH by utilising two novel algorithms; CCTP and VE. The CCTP algorithm produces filtered 4D Radar measurements in two different levels with low computational cost such that the resulting two measurements have different statistical characteristics, which enrich the input to the 4D Radar object detection network and leads to a significant performance improvement in the 3D object detection. The VE algorithm is an efficient vertical feature encoder that improves object detection performance with low computational cost and low memory requirement. We have demonstrated that the proposed RTNH+ achieves a significant performance improvement in 3D object detection with 4D Radar measurements of 10.14\% in ${AP}_{3D}^{IoU=0.3}$ and 16.12\% in ${AP}_{3D}^{IoU=0.5}$ over RTNH. In addition, we have tested various realizations of RTNH+ and verifies that the proposed RTNH+ achieves the best among all different realisations of CCTP and VE algorithms to RTNH.

\newcounter{appdxTableint}
\newcounter{appdxFigureint}
\newcommand\tabcounterint{%
  \refstepcounter{appdxTableint}%
  \renewcommand{\thetable}{\arabic{appdxTableint}}%
}
\newcommand\figcounterint{%
  \refstepcounter{appdxFigureint}%
  \renewcommand{\thefigure}{\arabic{appdxFigureint}}%
}
\setcounter{appdxFigureint}{4}
\setcounter{appdxTableint}{2}

\appendix

\newpage

\noindent\textbf{Apppendix} The appendix is organized as follows. We present detailed structure of RTNH++ and revised labels for K-Radar in Section A and B, respectively.
We discuss results for various hyperparameters of the combined CFAR-based two-level preprocessing (CCTP) and vertical encoding (VE) algorithms in Section D and E, respectively.
Finally, we show additional examples of 3D object detection using RTNH++ in Section E.

\section{Details of RTNH++}

In this subsection, we provide illustrations of the stage-1 CCTP and the stage-3 Neck \& Head of the proposed network, RTNH++.

\subsection{Illustration of the CCTP Algorithm}

\begin{figure}[h]
{
  \figcounterint
  \centering
  \includegraphics[width=1.0\columnwidth]{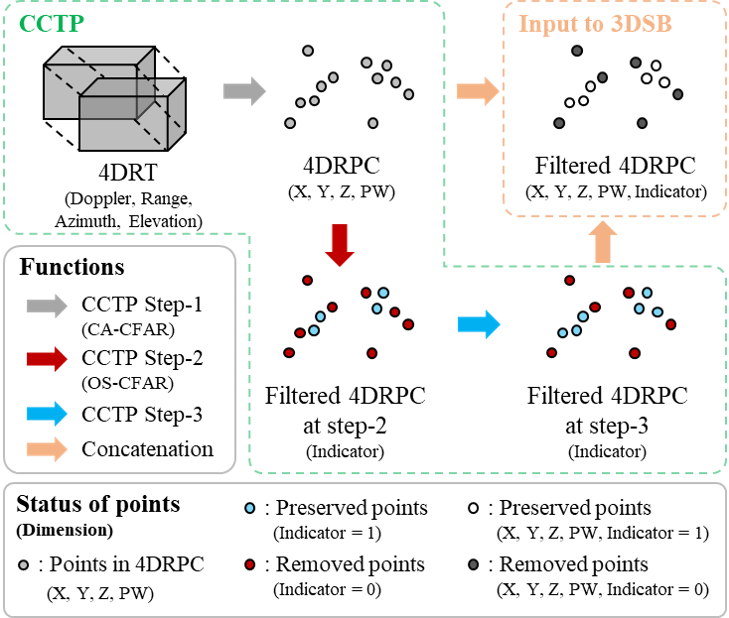}
  \caption{Combined CFAR-based Two-level Preprocessing (CCTP) of RTNH++.}
  \label{fig:cctp}
}
\end{figure}

As shown in Fig.\ref{fig:cctp}, CCTP step-1 applies a coarse CA-CFAR with a high CFAR $K_1$\% to the entire measurement 4DRT (i.e., the 4D Radar tensor) and extracts a 4D Radar point cloud (4DRPC) where measurements with very low power (e.g., background noise) are removed. Note that we do not consider Doppler measurements in stage-1, so each measurement (i.e., point) in the 4DRPC has values of X, Y, Z, and power. In CCTP step-2, we apply an OS-CFAR with a CFAR $K_2$\% to 4DRPC along the azimuth axis for each range, and some of the removed measurements in CCTP step-2 are restored in step-3. The resulting slightly-restored 4DRPC is used to generate indicators of 1 and 0 for the preserved measurements (of the filtered 4DRPC) and the removed measurements, respectively. As mentioned in subsection 3.2, some valid measurements may be lost and not included in the final output of CCTP step-3, which may cause higher miss-detection probability. To prevent miss-detection, we concatenate the output of CCTP step-1 (4DRPC) and the indicator output of CCTP step-3 to build an input to the 3DSB, where the 4DRPC is used as the primary input to the 3DSB to encode features, and the indicator is utilized for reliable information to precise object localization in the 3DSB. 

\subsection{Details of Neck \& Head}

\begin{figure}[h]
{
  \figcounterint
  \centering
  \includegraphics[width=1.0\columnwidth]{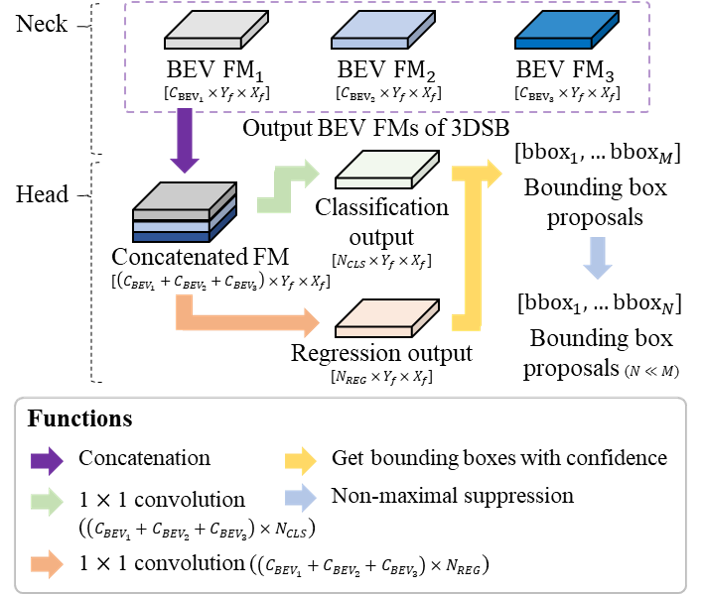}
  \caption{{Neck \& Head of RTNH++.}}
  \label{fig:n_and_h}
}
\end{figure}


As mentioned in subsection 3.2, we use the stage-3 Neck \& Head of RTNH \cite{paek1} for that of RTNH++. As shown in Fig.\ref{fig:n_and_h}, the Neck concatenates bird-eye-view (BEV) feature maps (FMs) that is generated in the stage-2 3DSB. The concatenated FM has a size of $(C_{{BEV}_1}+C_{{BEV}_2}+C_{{BEV}_3})\times{Y_{f}}\times{X_{f}}$, where ${C_{{BEV}_n}}$ represents the channel size of the n-th BEV FM and ${Y_{f}}$ and ${X_{f}}$ represent the size in X and Y axes of the 1st FM in the 3DSB.

Head uses an anchor-based method to predict the bounding boxes from the concatenated FM, which is similar to the method used in Faster R-CNN \cite{ren}. 
We apply 1$\times$1 convolutions to the concatenated FM to extract classification and regression outputs for each grid. Two anchor boxes with yaw angles of 0$^{\circ}$ and 90$^{\circ}$ are used for each class, which results in $N_{CLS}=2$ (anchor)$+1$ (background)$=3$. For each anchor, a total of eight parameters are assigned, which includes the center point ($x_c, y_c, z_c$), length, width, height ($x_l, y_l, z_l$), $\cos(yaw)$, and $\sin(yaw)$ of a bounding box, resulting in $N_{REG}=8$ (regression value)$\times2$ (anchor)$=16$. Then, we extract $M$ proposed bounding boxes from the classification and regression results.

During training, proposals with an intersection over union (IoU) of more than 0.5 with the ground-truth are classified as positive bounding boxes, and proposals with an IoU of less than 0.2 are classified as negative bounding boxes. To address the issue of class imbalance between positive and negative bounding boxes, we apply the focal loss \cite{lin}. The smooth L1 loss is used between the regression value and the ground-truth target value.

During inference, we infer the class of the proposal by selecting the index with the largest logit value from the classification output. We apply a confidence threshold of 0.3 to filter out low-confidence predictions, which are regarded as backgrounds. We then perform non-maximal suppression to remove overlapping bounding boxes, resulting in a total of $N$ bounding boxes.

\section{Details of revised labels for K-Radar}

\begin{figure}[h]
{
  \figcounterint
  \centering
  \includegraphics[width=1.0\columnwidth]{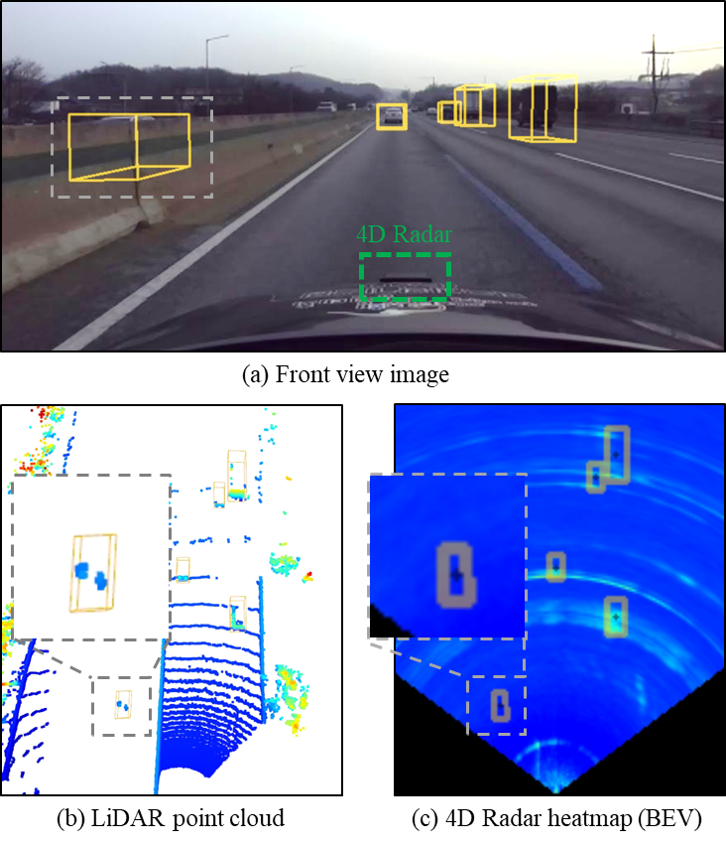}
  \caption{An example of a label of an object invisible to 4D Radar but visible to LiDAR, which is illustrated with gray dashed boxes in (a) the front view image, (b) LiDAR point cloud, and (c) 4D Radar heatmap.}
  \label{fig:label}
}
\end{figure}

\begin{table*}[t]
{
\tabcounterint
\begin{center}
\begin{tabular}{c|c|cc|cc|c|cc}
\hline\hline
\multirow{2}{*}{Network} & Hyperparameters & \multicolumn{2}{c|}{IoU=0.3} & \multicolumn{2}{c|}{IoU=0.5} & Mem & \multicolumn{2}{|c}{FPS [Hz]} \\
\cline{3-6} \cline{8-9}
   &  of CCTP & ${AP}_{3D}$  & ${AP}_{BEV}$ & ${AP}_{3D}$ & ${AP}_{BEV}$ & [MB] & S1-3 & S2 \\
\hline
$N_{{2.5},a}$ & 2.5-Nan-Nan & 53.93 & 62.36 & 25.84 & 46.93 & \textbf{323} & \textbf{16.23} & \textbf{98.86}\\
$N_{{2.5},b}$ & 2.5-2.5-(2,1) & 53.75 & 61.62 & 26.19 & 46.47 & 323 & 9.96 & 98.50\\\hline
$N_{{5},a}$ & 5-Nan-Nan & 52.86 & 60.49 & 25.96 & 45.07 & 359 & 15.79 & 97.85\\
$N_{{5},b}$ & 5-2.5-(2,1) & 54.55 & 62.53 & 22.84 & 44.85 & 359 & 9.62 & 97.84\\
$N_{{5},c}$ & 5-5-(2,1) & 53.83 & \textbf{63.02} & 26.36 & 46.31 & 359 & 8.91 & 97.32\\
$N_{{5},d}$ & 5-10-(2,1) & \textbf{54.77} & 62.57 & 24.68 & 47.25 & 359 & 8.30 & 97.50\\
$N_{{5},e}$ & 5-15-(2,1) & 53.28 & 61.19 & 27.30 & 45.80 & 359 & 5.99 & 97.12 \\\hline
$N_{{10},a}$ & 10-Nan-Nan & 50.66 & 58.94 & 25.85 & 44.72 & 387 & 14.95 & 94.55\\
$N_{{10},b}$ & 10-2.5-(2,1) & 54.22 & 62.33 & \textbf{27.90} & \textbf{50.39} & 387 & 6.61 & 94.43\\
$N_{{10},c}$ & 10-5-(2,1) & 53.81 & 61.34 & 27.17 & 47.19 & 387 & 5.62 & 94.58\\
$N_{{10},d}$ & 10-10-(2,1) & 54.06 & 61.76 & 25.53 & 46.87 & 387 & 4.90 & 93.96\\
$N_{{10},e}$ & 10-15-(2,1) & 51.29 & 58.44 & 26.77 & 44.85 & 387 & 4.56 & 94.15\\\hline
$N_{{20},a}$ & 20-Nan-Nan & 53.04 & 58.24 & 25.33 & 45.22 & 462 & 14.46 & 85.70\\
\hline
\hline
\end{tabular}
\end{center}
\caption{Performance comparison of networks with various hyperparameters of CCTP. The hyperparameters represent $K_1$-$K_2$-$(d_r, d_a)$ where $K1$, $K2$, and $(d_r, d_a)$ are the first and second step thresholds, and allowed distance in the range and azimuth axes in the CCTP, respectively. `Mem' and `FPS' represent the maximum GPU memory and computational cost required for inference in mega-bytes [MB] and frames per second (FPS), respectively. `S1-3' and `S2' represent  the FPS for the whole stages and for the stage-2 only, respectively. Note that the various networks we compare are denoted by ${N_{K1,a}}\sim{N_{K1,e}}$. The networks with the same $K1$ show similar performance in `Mem' and FPS of `S2', since the networks use the common primary input (i.e., output of CCTP step-1). In addition, the networks with the same $K1$ but larger $K2$ and those with the same $K2$ but larger $K1$ show lower performance in FPS of `S1-3', since denser measurements require larger computation cost for CCTP.}
\label{tab:cctp}
}
\end{table*}

\begin{table*}[t]
{
\tabcounterint
\begin{center}
\begin{tabular}{c|c|cc|cc|c|cc}
\hline\hline
\multirow{2}{*}{Network} & \multirow{2}{*}{$N_{h}$} & \multicolumn{2}{c|}{IoU=0.3} & \multicolumn{2}{c|}{IoU=0.5} & Mem & \multicolumn{2}{|c}{FPS [Hz]} \\
\cline{3-6} \cline{8-9}
   & & ${AP}_{3D}$  & ${AP}_{BEV}$ & ${AP}_{3D}$ & ${AP}_{BEV}$ & [MB] & S1-3 & S2 \\
\hline
$N_{1}$ & 1 & 52.37 & 63.74 & 29.48 & \textbf{55.67} & \textbf{501} & 9.14 & 105.1\\
$N_{2}$ (proposed) & 2 & \textbf{57.58} & \textbf{65.69} & \textbf{31.72} & 54.66 & 502 & \textbf{9.46} & \textbf{107.5}\\
$N_{4}$ & 4 & 54.84 & 62.38 & 30.25 & 51.70 & 504 & 8.99 & 104.3\\
$N_{8}$ & 8 & 54.55 & 62.53 & 22.84 & 44.85 & 510 & 8.89 & 103.7\\
\hline
\hline
\end{tabular}
\end{center}
\caption{Performance comparison of networks with various hyperparameters of VE. Note that the various networks we compare are denoted by ${N_{N_h}}$, where $N_h$ is the number of heads for multi-head attention in VE.}
\label{tab:ve}
}
\end{table*}

As outlined in subsections 2.2 and 3.3, most 4D Radar object detection datasets \cite{paek1, meyer, zheng} use LiDAR-based 3D bounding boxes as reference labels. However, a key difference between LiDAR and 4D Radar sensors is often their respective installation locations. In other words, while LiDAR is typically installed on the rooftop of a vehicle, 4D Radar is placed in front of or directly above the vehicle front bumper, as illustrated by the dashed green box in Fig.\ref{fig:label}-(a). As a result of this sensor placement disparity, objects detected by a LiDAR maybe not visible to a 4D Radar as shown in Fig.\ref{fig:label}-(b,c). This issue leads to the presence of labels in 4D Radar datasets belong to objects that are invisible to the Radar but visible to the LiDAR. This can negatively impact the performance of 4D Radar object detection neural networks, resulting in non-nelgigible increase of false alarms and detection performance degradation. To address this issue, we distinguish labels that are invisible to Radar in K-Radar and we demonstrate this revision can improve the performance of the 4D Radar object detection network as shown in Table 2.

\section{Experiments on various hyperparameters of CCTP}

Table \ref{tab:cctp} presents the quantitative 3D object detection performance of RTNH++ with various hyperparameters of CCTP. To compare the performance for various hyperparameters of the CCTP, we keep the configurations of stage-2 (encoding vertical features into a single value \cite{paek1,zhou}) and stage-3 (refer to subsection A.2) consistent for all networks, while only varying the hyperparameters of stage-1 (CCTP).

As mentioned in subsection 4.3, we discuss that small drop in PRVM may cause some (not large) drop in the detection probability of RTNH++, whereas a small drop in PRIM can cause a large drop due to a huge increase in the false alarm probability of RTNH++. This is because the number of invalid measurements is significantly higher than the number of valid measurements in the typical 4D Radar tensor measurements (4DRT), which is also observed in the K-Radar.

As shown in Table \ref{tab:cctp}, $N_{2.5,b}$ that has the lowest performance in PRVM (refer to Fig.3) shows only 0.08\%, 1.4\%, and 0.17\% lower performance in ${AP}_{3D}^{IoU=0.3}$, ${AP}_{BEV}^{IoU=0.3}$, and ${AP}_{3D}^{IoU=0.5}$, respectively, when compared to $N_{5,c}$ (i.e., the network with selected hyperparameters). On the other hand, $N_{10,a}$ and $N_{10,e}$ that has low performance in RRIM show significantly lower performance of 3.17\%, 4.08\%, and 1.59\% and 2.54\%, 4.58\%, and 1.46\% for ${AP}_{3D}^{IoU=0.3}$, ${AP}_{BEV}^{IoU=0.3}$, and ${AP}_{BEV}^{IoU=0.5}$, respectively, when compared to $N_{5,c}$. Therefore, we consider hyperparameters that lead to moderate performance in PRVM and high performance in RRIM, which are 5-5-(2,1) (i.e., the hyperparameters of $N_{5,c}$) or 10-2.5-(2,1) (i.e., the hyperparameters of $N_{10,b}$). As shown in Table \ref{tab:cctp}, $N_{5,c}$ shows the best performance in ${AP}_{BEV}^{IoU=0.3}$ and $N_{10,b}$ results in the best performance in 
${AP}_{3D}^{IoU=0.5}$ and ${AP}_{BEV}^{IoU=0.5}$. However, considering the computational cost, we select 5-5-(2,1) as the final hyperparameter for CCTP.

\section{Experiments on various hyperparameters of VE}

Table \ref{tab:ve} presents the quantitative 3D object detection performance of RTNH++ for various hyperparameters of VE. To compare the performance for various  hyperparameters of VE, we keep the configuration of stage-1 (CCTP) and stage-3 (refer to subsection A.2) consistent for all networks, while only varying the number of heads for multi-head attention (MHA) in VE.

As mentioned in subsection 3.2, we employ MHA \cite{vaswani}, which has an advantage in computation speed due to the parallel encoding, for correlation between vertical query and features. As shown in Table \ref{tab:ve}, we observe that the networks with 4 or more heads ($N_4$, $N_8$) do not have any enhancement in both $AP$ and FPS but require more GPU memory. Therefore, we propose the network with 2 heads ($N_2$), which leads to the best performance in ${AP}_{3D}^{IoU=0.3}$, ${AP}_{BEV}^{IoU=0.3}$, ${AP}_{3D}^{IoU=0.5}$, and FPS.

\begin{figure*}[ht]
{
  \figcounterint
  \centering
  \includegraphics[width=2.0\columnwidth]{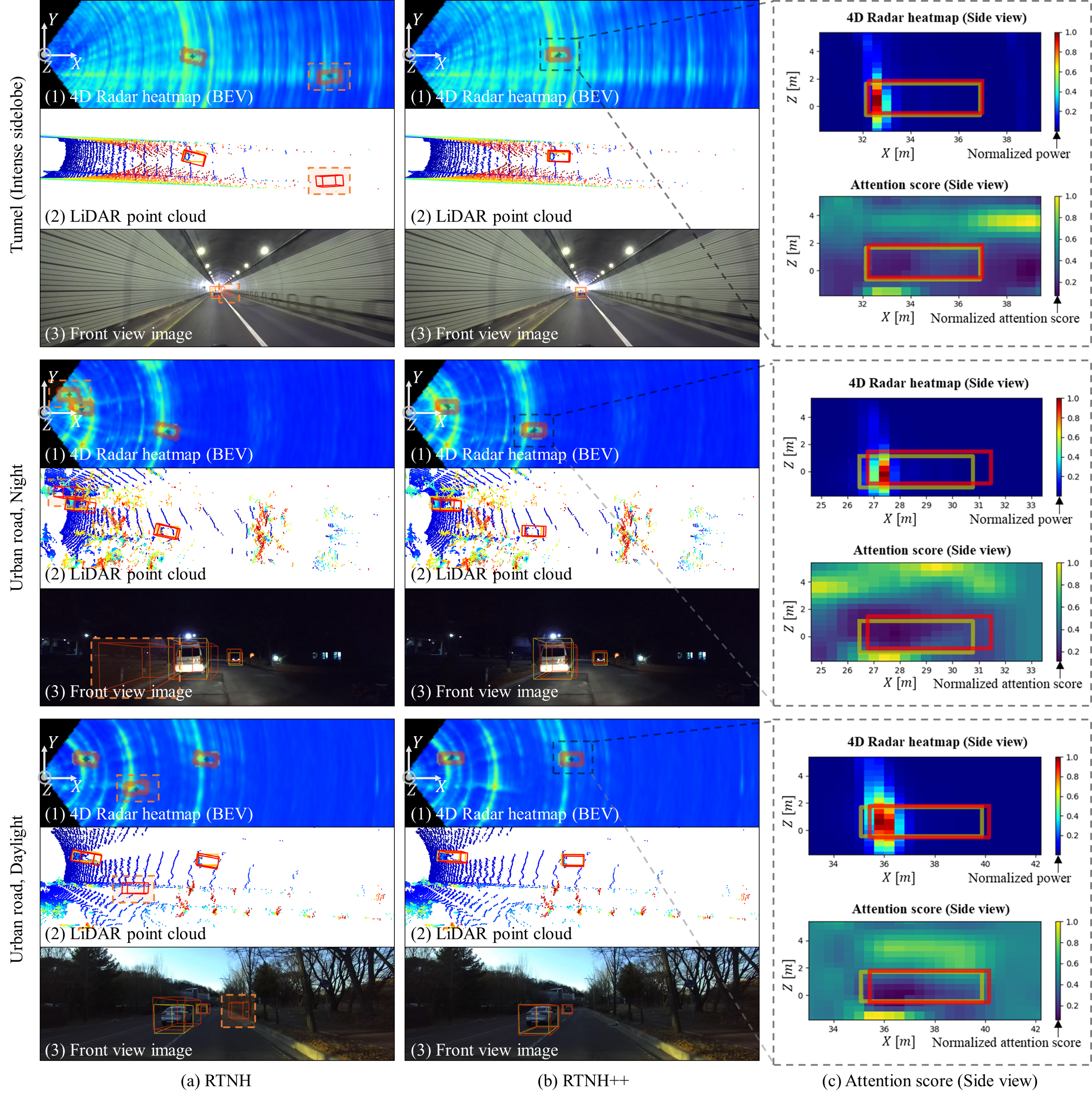}
  \caption{Examples of 3D object detection using (a) RTNH and (b) RTNH++ for three different road environments described on the left of (a). Each cell of (a) and (b) show three figures in a column that (1) 4D Radar heatmap in BEV, (2) LiDAR point cloud, and (3) front view image with bounding boxes, at the same site. Bounding boxes with yellow solid line, red solid line, and orange dashed line represent the ground truths, predictions, and false alarms, respectively. The third column (c) shows the 4D Radar heatmap of the object and attention score of vertical query on a side-view (on the XZ-plane) for the three environments.}
  \label{fig:infer_1}
}
\end{figure*}

\begin{figure*}[ht]
{
  \figcounterint
  \centering
  \includegraphics[width=2.0\columnwidth]{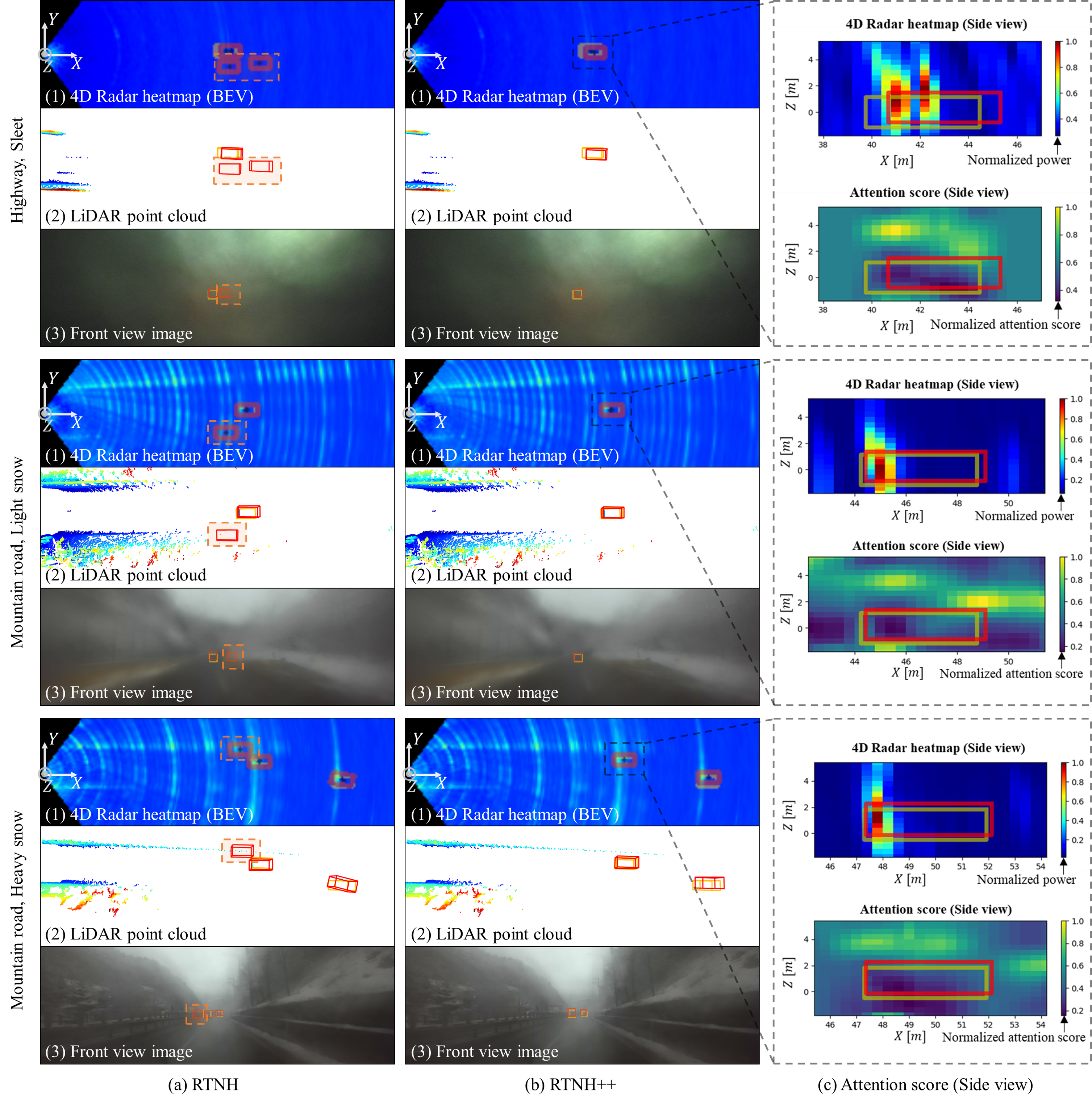}
  \caption{Examples of 3D object detection using (a) RTNH and (b) RTNH++ for three different road environments under different weather conditions with precipitation described on the left of (a). Each cell of (a) and (b) show three figures in a column that (1) 4D Radar heatmap in BEV, (2) LiDAR point cloud, and (3) front view image with bounding boxes, at the same site. Bounding boxes with yellow solid line, red solid line, and orange dashed line represent the ground truths, predictions, and false alarms, respectively. The third column (c) shows the 4D Radar heatmap of the object and attention score of vertical query on a side-view (on the XZ-plane) for the three environments.}
  \label{fig:infer_2}
}
\end{figure*}

\section{Qualitative results of RTNH and RTNH++ in various road environments}

The following Fig.\ref{fig:infer_1} and \ref{fig:infer_2} show the qualitative 3D object detection results of RTNH and RTNH++ in various road environments, including adverse weather conditions. Note that Fig.\ref{fig:infer_2} shows the robustness of 4D Radar to adverse weather conditions; even in road environments where camera and LiDAR measurements for objects are lost due to the heavy snow or sleet, 4D Radar can provide measurements for objects. And then the 3D object detection networks such as RTNH or RTNH++ can detect objects based on the 4D Radar measurements.

In subsection 4.3, both Table 1 and Fig.4 demonstrate that RTNH++ shows (1) higher object detection performance due to both lower miss-detection and lower false alarm, (2) more precise localization performance in both BEV and 3D by utilizing proposed CCTP and VE, resulting in 10.14\%, 7.3\%, 11.48\%, and 16.12\% increase in ${AP}_{3D}^{IoU=0.3}$, ${AP}_{BEV}^{IoU=0.3}$, ${AP}_{3D}^{IoU=0.5}$, and ${AP}_{BEV}^{IoU=0.5}$, respectively, over RTNH. Similarly, both Fig.\ref{fig:infer_1} and \ref{fig:infer_2} demonstrate the superiority of CCTP and VE that constitute RTNH++. RTNH in 1st, 2nd, and 3rd rows of both Fig.\ref{fig:infer_1} and \ref{fig:infer_2} show examples of false alarms. In particular, as shown in 4D Radar heatmap in the 2nd row of Fig.\ref{fig:infer_1} and 1st and 3rd rows of Fig.\ref{fig:infer_2}, RTNH falsely detects objects near the ground-truth objects due to the strong sidelobes along the azimuth axis. On the other hand, RTNH++ does not generate false alarms due to the efficient filtering of invalid measurements by the CCTP, as shown in Fig.\ref{fig:infer_1}-(b) and \ref{fig:infer_2}-(b). Fig.\ref{fig:infer_1}-(c) and \ref{fig:infer_2}-(c) show the 4D Radar heatmap and attention score of the vertical query on a side-view (XZ-plane). As discussed in subsection 4.3, the attention score shows that the vertical query encodes useful vertical features for 3D object detection, such as the vertical boundaries of the 3D bounding boxes, rather than the entire vertical features.

{\small
\bibliographystyle{ieee_fullname}
\bibliography{egbib}
}

\end{document}